\documentclass[a4paper,11pt]{article}
\usepackage{jinstpub} 
\usepackage{siunitx}
\usepackage{subcaption}
\usepackage{hyperref}
\newcommand{\cevns}{{\text{CE}$\nu$\text{NS}}}

\usepackage{enumitem}

\DeclareSIUnit{\PE}{PE}
\DeclareSIUnit{\sample}{\text{S}}
\DeclareSIUnit{\faraday}{\text{F}}
\DeclareSIUnit{\dB}{\text{dB}}
\DeclareSIUnit{\event}{\text{event}}

\title{\boldmath Design and characterization of a photosensor system for the RELICS experiment}

\author[a,b,2]{Jijun~Yang,%
\note[2]{Corresponding author.}}
\author[a,3]{Ruize~Li,%
\note[3]{Now at: Cavendish Laboratory, University of Cambridge, Cambridge, CB3 0US, UK.}}
\author[c]{Chang~Cai,}
\author[d]{Guocai~Chen,}
\author[e]{Jiangyu~Chen,}
\author[f]{Huayu~Dai,}
\author[g]{Rundong~Fang,}
\author[c]{Fei~Gao,}
\author[c]{Jingfan~Gu,}
\author[h,i]{Xiaoran~Guo,}
\author[g]{Jiheng~Guo,}
\author[d]{Gaojun~Jin,}
\author[d]{Fali~Ju,}
\author[c]{Yanzhou~Hao,}
\author[c]{Yang~Lei,}
\author[c]{Kaihang~Li,}
\author[d]{Meng~Li,}
\author[d]{Minhua~Li,}
\author[a,2]{Shengchao~Li,}
\author[a]{Siyin~Li,}
\author[d]{Tao~Li,}
\author[h,i]{Qing~Lin,}
\author[j]{Jiajun~Liu,}
\author[d]{Sheng~Lv,}
\author[j]{Guang~Luo,}
\author[c]{Kangwei~Ni,}
\author[d]{Chuanping~Shen,}
\author[c]{Mingzhuo~Song,}
\author[h,i]{Lijun~Tong,}
\author[a,b]{Jun~Wang,}
\author[a]{Xiaoyu~Wang,}
\author[e,j]{Wei~Wang,}
\author[g,k]{Xiaoping~Wang,}
\author[d]{Zihu~Wang,}
\author[f]{Yuehuan~Wei,}
\author[d]{Liming~Weng,}
\author[j]{Xiang~Xiao,}
\author[c]{Lingfeng~Xie,}
\author[l]{Litao~Yang,}
\author[d]{Long~Yang,}
\author[f]{Jingqiang~Ye,}
\author[h,i]{Jiachen~Yu,}
\author[l]{Qian~Yue,}
\author[e]{Yuyong~Yue,}
\author[d]{Bingwei~Zhang,}
\author[c]{Yuming~Zhang,}
\author[c]{Yifei~Zhao}
\author[i]{and Chenhui~Zhu}

\affiliation[a]{School of Science, Westlake University, Hangzhou 310030, China}
\affiliation[b]{Institute of Natural Sciences, Westlake Institute for Advanced Study, Hangzhou 310024, China.}
\affiliation[c]{Department of Physics \& Center for High Energy Physics, Tsinghua University, Beijing 100084, China}
\affiliation[d]{CNNC Sanmen Nuclear Power Company, Zhejiang 317112, China}
\affiliation[e]{Sino-French Institute of Nuclear Engineering and Technology, Sun Yat-sen University, Zhuhai 519082, China}
\affiliation[f]{School of Science and Engineering, The Chinese University of Hong Kong (Shenzhen), Shenzhen, Guangdong, 518172, China}
\affiliation[g]{School of Physics, Beihang University, Beijing 100083, China}
\affiliation[h]{State Key Laboratory of Particle Detection and Electronics, University of Science and Technology of China, Hefei 230026, China}
\affiliation[i]{Department of Modern Physics, University of Science and Technology of China, Hefei 230026, China}
\affiliation[j]{School of Physics, Sun Yat-sen University, Guangzhou 510275, China}
\affiliation[k]{Beijing Key Laboratory of Advanced Nuclear Materials and Physics, Beihang University, Beijing 100191, China}
\affiliation[l]{Key Laboratory of Particle and Radiation Imaging (Ministry of Education) \& Department of Engineering Physics, Tsinghua University, Beijing 100084, China}

\collaboration{RELICS Collaboration\footnote{Collaboration email:
\href{mailto:relics@tsinghua.edu.cn}{relics@tsinghua.edu.cn}}}

\emailAdd{yangjijun@westlake.edu.cn}
\emailAdd{lishengchao@westlake.edu.cn}

\abstract{
In this paper, we present the design and characterization of a photosensor system developed for the RELICS experiment. An extended dynamic range base was designed to mitigate photomultiplier tube (PMT) saturation caused by intense cosmic muon backgrounds in the surface-level RELICS detector. The system employs dual readout from the anode and the seventh dynode to extend the linear response range of the PMT. In particular, our characterization and measurements of Hamamatsu R8520-406 PMTs confirm stable operation under positive high-voltage bias, extending the linear response range by more than an order of magnitude. Furthermore, a model of PMT saturation and recovery was developed to evaluate the influence of cosmic muon signals in the RELICS detector. The results demonstrate the system capability to detect coherent elastic neutrino–nucleus scattering signals under surface-level cosmic backgrounds, and suggest the potential to extend the scientific reach of RELICS to MeV-scale interactions.
}


\arxivnumber{2510.24196} 

\begin{document}
\maketitle
\flushbottom

\section{Introduction}\label{sec:intro}
The detection of coherent elastic neutrino-nucleus scattering (\cevns) is not only crucial for constraining the standard model (SM) of particle physics~\cite{Freedman:1973yd, Kopeliovich:1974mv}, but it also provides new experimental techniques and theoretical frameworks that advance interdisciplinary research across astrophysics, nuclear physics, and dark matter detection. For example, \cevns\ is crucial for measuring the weak mixing angle in the SM at low energy~\cite{Canas:2018rng}, investigating non-standard neutrino interactions~\cite{Barranco:2007tz}, searching for sterile neutrinos~\cite{Anderson:2012pn}, and exploring neutrino electromagnetic properties~\cite{Dodd:1991ni}. Furthermore, \cevns\ offers a unique window into the study of nuclear structure~\cite{Patton:2012jr}, and plays a critical role in understanding the mechanisms driving core-collapse supernova explosions~\cite{Suliga:2021hek}. The \cevns\ signals originating from solar and atmospheric neutrinos constitute an irreducible background in dark matter detection experiments~\cite{XENON:2024ijk, PandaX:2024muv}. Thus, the precise measurement of \cevns\ is essential for dark matter searches beyond the so-called ``neutrino floor''~\cite {Billard:2013qya, OHare:2021utq}, enhancing the experimental sensitivity of those experiments. 

The dual-phase liquid xenon (LXe) time projection chamber (TPC) has been the leading technology in the direct detection of dark matter for decades~\cite{Aprile:2009dv, Chepel:2012sj, XENON100:2011cza, LUX:2016ggv, PandaX:2014ria}. The low energy threshold of ionisation signals makes LXe TPC particularly advantageous for detecting weak signals such as \cevns. Using a cylindrical TPC of $\SI{24}{\centi\meter}$ in height and $\SI{28}{\centi\meter}$ in diameter, with arrays of $64$ Hamamatsu R8520-406 1-inch photomultiplier tubes (PMTs) both at top and bottom, the REactor neutrino LIquid xenon Coherent Scattering (RELICS) experiment aims to take advantage of this robust, low-threshold technology to detect \cevns\ process of reactor neutrinos at the Sanmen nuclear reactor site~\cite{RELICS:2024opj}. Unlike the situation in deep-buried dark matter experiments, the conducting of surface-level low-energy event searches presents significant challenges
due to cosmic-ray related background~\cite{COHERENT:2025vuz, TEXONO:2024hnv, CONNIE:2024pwt, Chattaraj:2025fvx, RED-100:2024izi}. Among these, cosmic muons are one of the main sources of background in the RELICS experiment, at the rate of $\SI{1}{\event\per\centi\meter\squared\per\minute}$~\cite{Autran:2018etp}. 

\begin{figure}[htbp]
    \centering
    \includegraphics[width=0.7\linewidth]{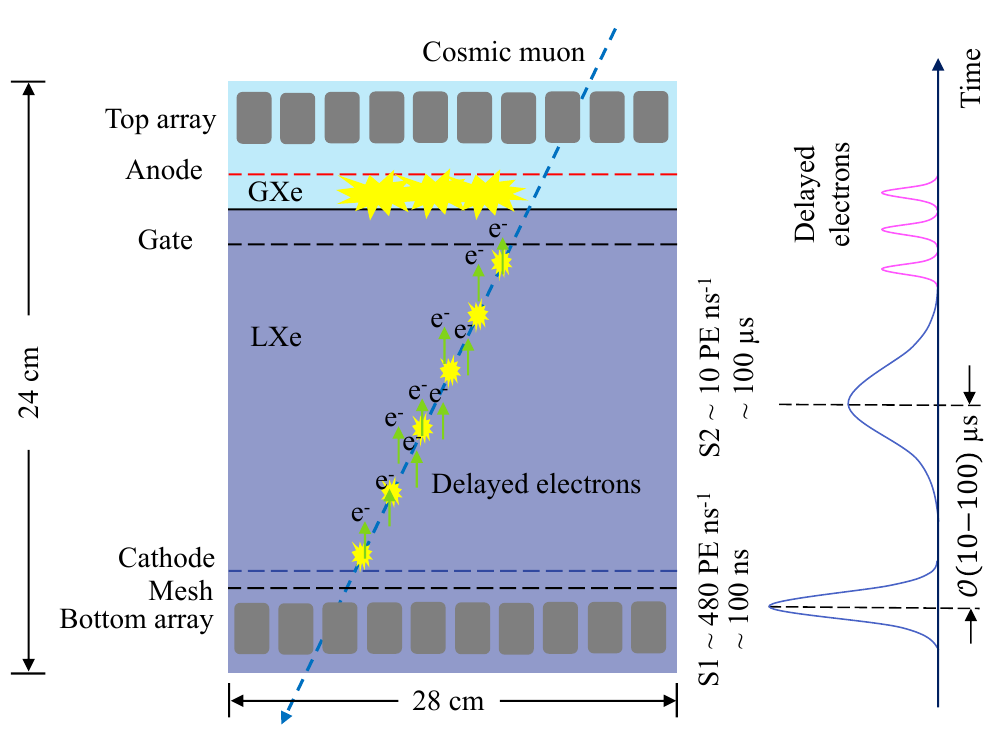}
    \caption{The schematic diagram of a cosmic-ray muon traversing through the RELICS LXe TPC, producing the initial S1 and S2 signals (diagram not to scale). Delayed emission of single- and few-electron follows the initial ionization signal by muons, with a $\SI{0.174}{\centi\meter\per\micro\second}$ electron drift velocity~\cite{RED-100:2024izi} under a $\SI{500}{\volt\per\centi\meter}$ drift field, which shows both position and time correlations. }
    \label{fig:muon_track}
\end{figure}

The energy deposition of muons will cause strong prompt scintillation (S1) and produce a large number of ionisation electrons responsible for the secondary proportional ionisation scintillation (S2), as shown in figure~\ref{fig:muon_track}. Both S1 and S2 from muons will cause the PMT anodes to saturate, but to different extents. This saturation effect is known to have time and energy dependences as the PMTs recharge and recover. Moreover, the muon-induced backgrounds, such as neutrons and delayed electrons, are the key background components for the search of \cevns~\cite{XENON:2021qze}, which requires a precise reconstruction of the muon directional information~\cite{RELICS:2024opj}. Both effects are crucial for the scientific goals of the RELICS experiment, which require the photo-sensor system to have an extended dynamic range to the higher energy side. 

We organize our paper as follows: We begin by analysing the muon-related background in the RELICS experiment, outlining the motivations and requirements for dynamic readout. Then we describe the design of the extended dynamic range base, especially the design of the dynode readout in section~\ref{sec:design}. The performance and tests of the dynode (anode) readout are discussed in section~\ref{sec:dp} (\ref{sec:anode}). Based on these results, we investigate the saturation responses of PMTs and model the time- and energy-dependent saturation behavior, laying a foundation for a full signal recovery technology. Finally, we close the topic by discussing the current design and possible outlooks in section~\ref{sec:discussion}. 

\section{Muon-induced backgrounds}\label{sec:muon-bg}
The RELICS detector is located at the surface, where no natural shielding exists. As a result, it is directly exposed to cosmic muons with a mean energy of approximately $\SI{4}{\giga\electronvolt}$ and event rate $\mathcal{O}(\SI{10}{})~\SI{}{\hertz}$, according to the size of the RELICS TPC~\cite{Nakamura_2010, Autran:2018etp}. This results in a significant background event rate, and an elevated requirement for data bandwidth. 

The RELICS experiment mainly focuses on the detection of the low-energy \cevns\ signal from $\SI{0.63}{\kilo\electronvolt}$ to $\SI{1.36}{\kilo\electronvolt}$, roughly equivalent to $120$ to $300$ photo-electrons (PE). This naturally requires the detector to be extremely sensitive to low-energy events~\cite{RELICS:2024opj}. 
However, an average cosmic-ray muon deposit O(1) MeV energy in the LXe, far exceeding this
range.
In RELICS TPC, a vertical muon could deposit $\SI{120}{\mega\electronvolt}$ inside the fiducial volume, producing $\SI{e6}{\PE}$ within $\SI{100}{\nano\second}$. The mean S1 intensity received by each PMT is therefore about $\SI{480}{\PE\per\nano\second}$~\cite{jason_brodsky_2019_2535713}. Such a large S1 signal can easily cause the saturation of PMTs. At the same time, the mean intensity of S2 can reach up to $\SI{10}{\PE\per\nano\second}$ with a long duration of up to $\SI{178}{\micro\second}$. 

Moreover, delayed ionization signals comprising only single or a few electrons are observed in liquid xenon TPCs after large S2 and can persist for up to $\mathcal{O}(1)$ second~\cite{XENON:2021qze, PandaX:2022xqx, LUX:2020vbj}. These delayed electrons can lead to substantial accidental coincidence backgrounds, adversely affecting \cevns\ detection. Research from XENON1T~\cite{XENON:2021qze} and LUX~\cite{LUX:2020vbj} indicates that delayed electrons exhibit correlations with physical processes in both time and space. In RELICS, the delayed electron pile-up background in the low-energy region could exceed the \cevns\ signal by four orders of magnitude~\cite{RELICS:2024opj}, which can be suppressed by considering the time and spatial correlations of delayed electrons with the muon trajectory. Thus, dynamic readout technologies need to be developed and optimized to address PMT saturation and enable the accurate collection of both energy and timing information from muon events. In addition, understanding of the muon trajectory would allow even precise tagging of the muon-induced neutron background, albeit in ref.~\cite{RELICS:2024opj}, such a background is deemed subdominant. 

A solution to this challenge is to employ a dedicated dynode readout scheme for the PMTs, which provides an inherently linear, low-gain signal alternative to the (potentially) saturated anode. Dynode readout has been widely used to precisely reconstruct high-energy events in other experiments~\cite{Zheng:2020kfp, Li:2021aom}. With carefully chosen parameters, dynode readout preserves both the timing information and the original waveform, enabling the accurate measurement of both the intense S1 and the otherwise affected S2 without distortion. This inherent capability makes it the optimal solution for photo-sensors in the RELICS experiment. 

\section{The design of the extended dynamic range base}\label{sec:design}

The front-end readout electronics for the RELICS experiment are designed to accommodate a wide signal range from low-energy \cevns\ events to high-energy cosmic-ray muons. As a measure of electron amplification, the gain $G$ of a PMT is determined by the configuration of the dynode voltages in the multiplier chain. The gain up to the $i$-th dynode (with $i=11$ corresponding to the anode in the Hamamatsu R8520-406), where the potential difference between the $j$-th dynode and its precedent dynode is $V_j$, can be approximated as
\begin{equation}
    G(i) \;=\; \prod_{j=1}^{i} g_j 
    \;=\; \prod_{j=1}^{i} K \, V_j^{\alpha} ,
    \label{eq:v-div}
\end{equation}
where $g_j$ is the secondary emission gain at the $j$-th dynode, $K$ is a constant depending on the dynode material and its geometrical structure, and $\alpha$ is an empirical parameter typically in the range $0.65 \lesssim \alpha \lesssim 0.75$~\cite{Flyckt:712713}. The overall PMT gain (at the anode) is then $G_{\text{total}} \;\equiv\; G(i=11)$.

\begin{figure}[htbp]
    \centering
    \includegraphics[width=\linewidth]{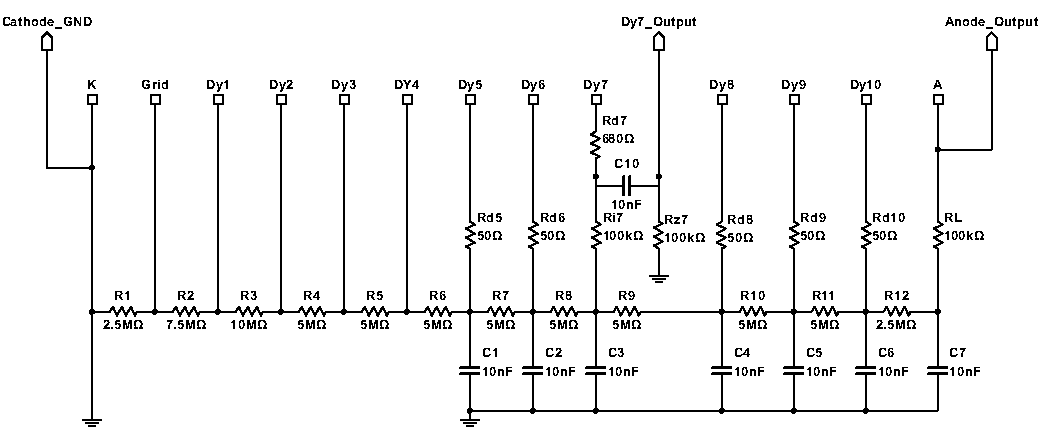}
    \caption{The dynamic readout PMT base circuit designed for the RELICS experiment. The RELICS PMTs are positively-biased with an input voltage of $\sim\SI{800}{\volt}$. A decoupler (see appendix~\ref{sec:decoupler}) is needed for the anode output (\texttt{Anode\_Output}) for a regular-gain signal; the low-gain signal is acquired from the seventh dynode (\texttt{Dy7\_Output}). }
    \label{fig:base_circuit}
\end{figure}

Based on our measurements, the total gain $G_{\text{total}}$ of the seven Hamamatsu R8520-406 PMTs ranges from $\SI{5.1e6}{}$ to $\SI{9.6e6}{}$, when the total operating voltage $V_\mathrm{total} = \Sigma_j V_j$, equals $\SI{800}{\volt}$. We determine the value of $K$ between $0.058$ and $0.061$ according to the official recommendation~\cite{Hamamatsu:pmt_manual}. Figure~\ref{fig:base_circuit} shows our design of the dynamic readout PMT base circuit. Voltage division between the cathode and the second dynode is set to approximately one-third of $V_\mathrm{total}$, maximizing photoelectron collection efficiency at the first dynode. 

A previous study~\cite{Zheng:2020kfp} reveals that for a PMT with a similar total gain, the anode saturation amplitude is about $\SI{1000}{\PE}$, about $\SI{2}{\percent}$ of the signal strength from a muon S1 in RELICS. Following this calculation and the scaling relation in eq.~\eqref{eq:v-div}, we choose to read out the seventh dynode for the $\mathcal{O}$(MeV) signals in RELICS, where $G(7)$ is expected to be suppressed by a factor of $(109\pm 8)$ compared to the anode, providing waveforms without saturation. A blocking capacitor ($C_{10}$) is used to decouple the AC signal, as shown in figure~\ref{fig:base_circuit}. 
The base circuit also provides a positive bias to the anode while keeping the photocathode and shell grounded. This design prevents the large high‑voltage gap between the top PMT array and the TPC anode, which can otherwise generate undesired ``gas events''~\cite{PandaX:2022xqx}, where the S1 and S2 signals merge and appear as background in S2-only analyses in the RELICS experiment. 

To stabilize the inter-dynode voltage and mitigate space charge effects~\cite{Zheng:2020kfp, Li:2021aom}, we employed six parallel capacitors ($C_{1}$\text{--}$C_{6}$, $\SI{10}{\nano\farad}$) connected to ground to extend the linear response range and maintain stable PMT performance. High-impedance resistors ($R_{i7}$ and $R_{L}$, $\SI{100}{\kilo\ohm}$) were added to suppress signal current backflow~\cite{Li:2021aom}, while the damping resistors ($R_{d5}$, $R_{d6}$, $R_{d8}$\text{--}$R_{d10}$, $\SI{50}{\ohm}$) were added to suppress the periodic discharge of the capacitors, known as the ``ringing effect''~\cite{Hamamatsu:pmt_manual}. Furthermore, internal parasitic inductance from the PMT is compensated by an isolated quenching resistor ($R_{d7}$, $\SI{680}{\ohm}$), whose value was optimized experimentally to suppress damped oscillations, by monitoring the falling edge of the dynode waveforms~\cite{Li:2021aom}.

\section{Dynode performance}\label{sec:dp}
\subsection{Bench test setup}

\begin{figure}[htbp]
    \centering
    \includegraphics[width=0.95\linewidth]{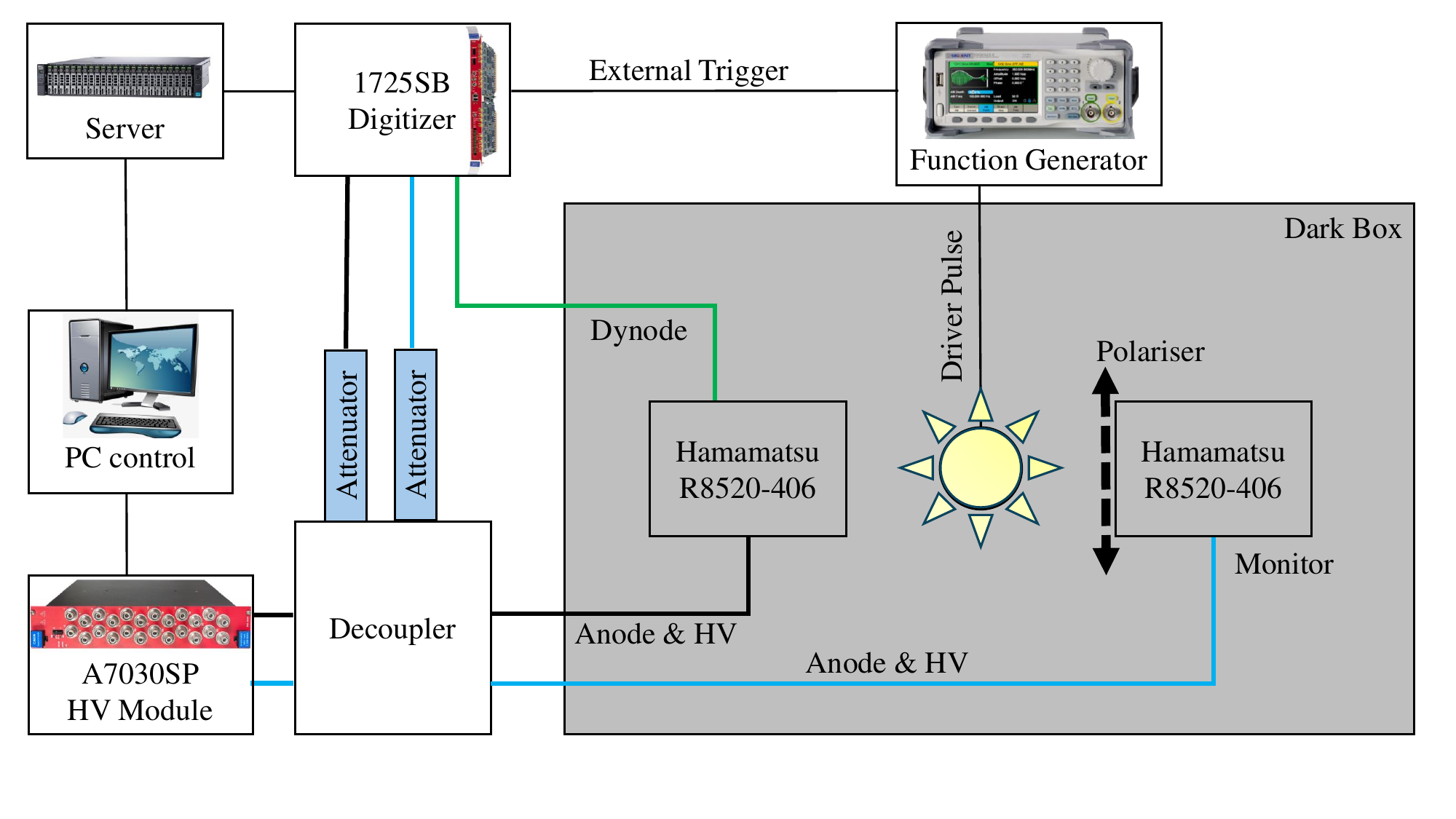}
    \caption{Box diagram of the bench test setup. The R8520-406 PMT with both anode and dynode readout is used to test the dynamic readout range of the base, while the other one is fitted with a light attenuator and used as an independent monitor. An LED, driven by a functional pulse generator (SIGLENT SDG6052X, $\SI{500}{\mega\hertz}$ bandwidth, $\SI{2.4}{GS \per\second}$ sampling rate), is placed between the PMTs and wrapped by a Teflon sphere to ensure the light emission is as uniformly distributed as possible over the entire $4\pi$ solid angle. The pulse generator driving the LED also provides synchronous triggers to a CAEN V1725SB digitizer ($\SI{250}{MS\per\second}$ sampling rate, 14-bit resolution, and $\SI{2}{\volt}$ dynamic range).
    }
    \label{fig:test_setup}
\end{figure}
\begin{figure}[htbp]
    \centering
    \includegraphics[width=9cm]{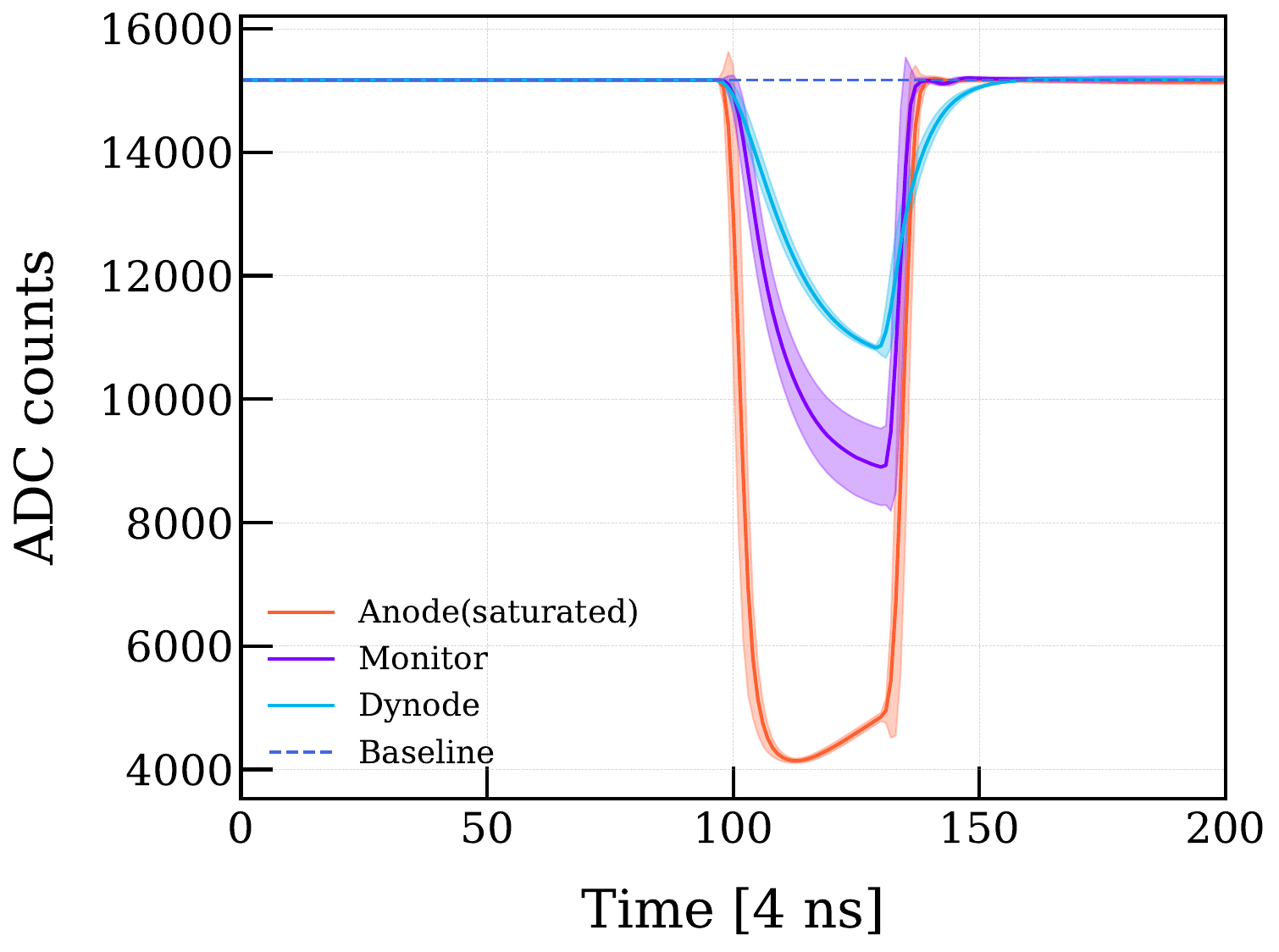}
    \caption{The waveforms shown include signals from the anode (light red), dynode (light blue), and monitor (light purple) PMT readout channels, recorded at an LED driver voltage of $\SI{1.58}{\volt}$. The shaded region represents a fluctuation of the waveform of each channel. The waveform from the dynode is reversed polarity for comparison with the anode and monitor waveforms. The horizontal axis is set in $\SI{4}{\nano\second}$ which is the time resolution of the ADC (this also applies to subsequent figures).}
    \label{fig:waveform_at_1.5V}  
\end{figure}

As depicted in figure~\ref{fig:test_setup}, the bench test apparatus is enclosed within a light-sealed box. A second PMT was employed to monitor the high-intensity light, with a polariser attenuating (factor of $83.0\pm 2.6$, measured independently) the intensity to the linear response range of the PMT. Both PMTs were supplied with positive $\SI{800}{\volt}$ high voltage by the CAEN A7030SP high voltage module.  In addition, to avoid the analog-to-digital converter (ADC) saturation, we applied two attenuators after the signals were decoupled from the high voltage. Figure~\ref{fig:waveform_at_1.5V} demonstrates the feature of the signal from the dynode, the anode of the test PMT, and the monitor PMT of our bench test. The anode signal from the test PMT was attenuated by $\SI{9}{dB}$ (about $\SI{11000}{ADC}$). The amplitude from the monitor PMT signal is approximately $\SI{6300}{ADC}$. The pulse shapes of the dynode are expected and exhibit good quality. Measurements using this setup are based on the relation between measured light intensity and the voltage applied to the LED, as shown in appendix~\ref{sec:appendix}. 

\begin{figure}[htbp]
    \centering
    \includegraphics[width=8cm]{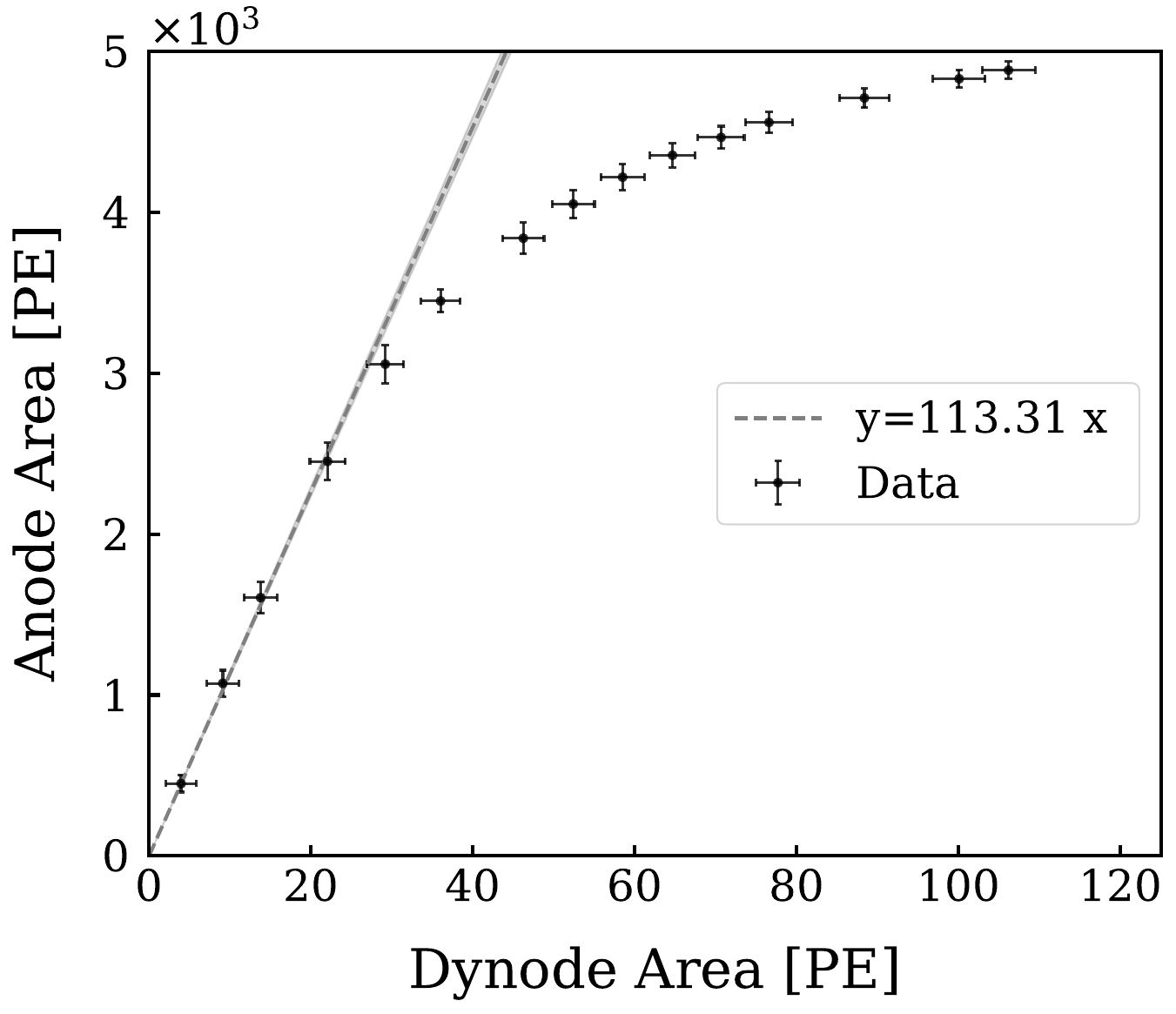}
    \caption{The anode output charge is shown as a function of the dynode readout charge, with three sigma error. The grey(shadow) dashed line indicates a linear fit (the error of the fitting), where the slope represents the gain ratio between the anode and dynode. At anode saturation, this ratio can become 43.39 using the last data point in the figure below. }
    \label{fig:adratio}  
\end{figure}

\subsection{Dynamic readout range}
\begin{figure}[htbp]
    \centering
    \includegraphics[width=9cm]{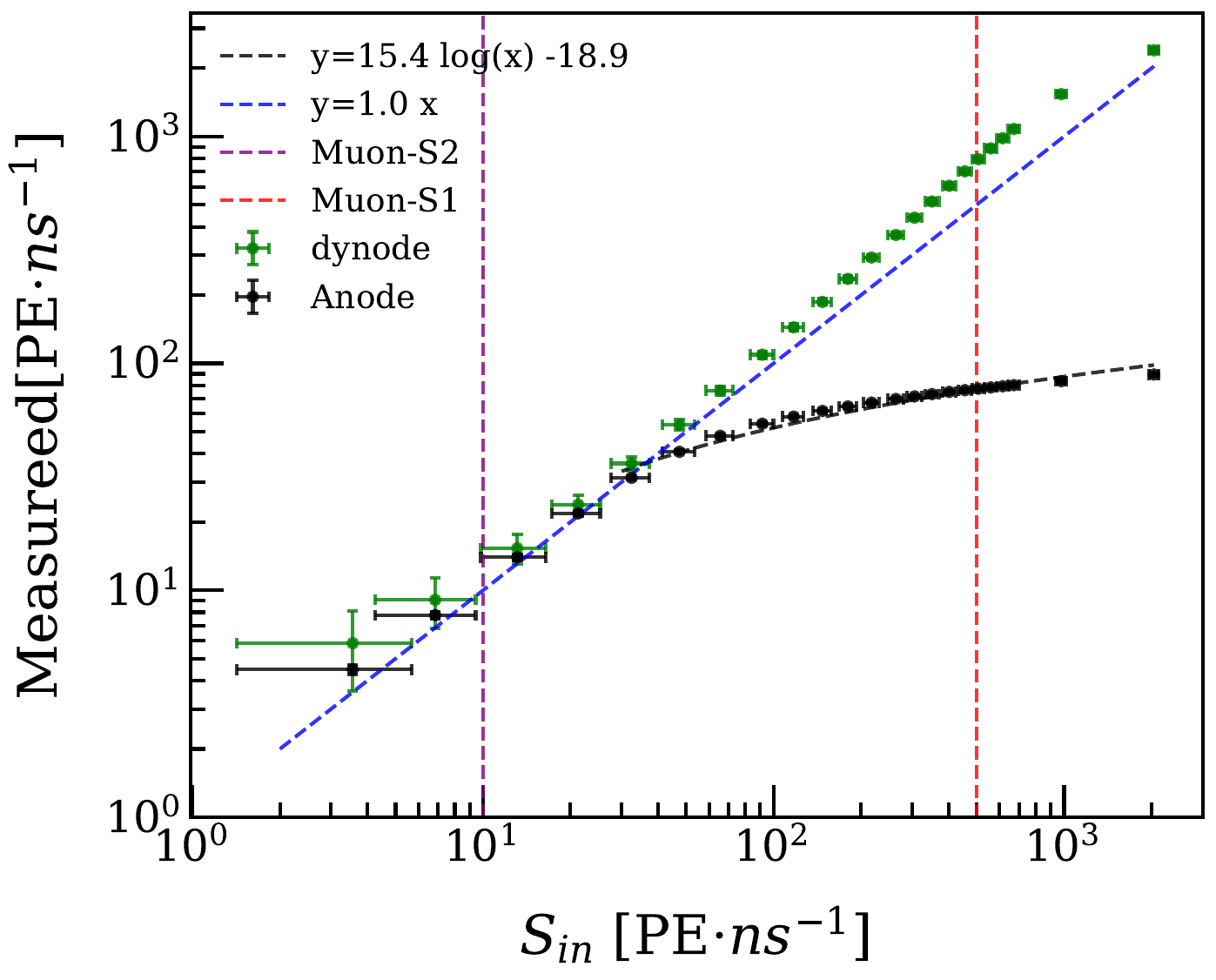}
    \caption{The dynamic readout range of R8520-406 and base under different incident light intensities. The green data points with errors are the readout light intensity from the dynode. The black data points are related to the anode, and the $x$-axis of incident light intensity is calibrated by the monitoring PMT. The red (purple) dashed line is related to the average charge density of muon S1 (S2). }
    \label{fig:dynamic_range}  
\end{figure}

We integrated the pulse area from every readout channel and fitted the charge spectrum with a Gaussian function for each run, and normalized to $\SI{}{\PE\per\nano\second}$ with the width of the LED pulse. Experimentally, the ratio from anode to dynode was confirmed to be $113.3$, shown in figure~\ref{fig:adratio}, which is consistent with the calculation in section~\ref{sec:design}. As demonstrated in figure~\ref{fig:dynamic_range}, the dynamic readout range of R8520-406 is expanded to over $\SI{1000}{\PE\per\nano\second}$ using dynode readout (green data points), exceeding the $\SI{500}{\PE\per\nano\second}$ (red dashed line) requirement for cosmic‐muon detection. The intensities from dynode readout are scaled by the anode-to-dynode gain ratio, as determined from test PMT measurements. The anode output saturates above roughly $\SI{40}{\PE\per\nano\second}$, which follows a natural logarithmic law. The black-dashed line represents the anode data fitted by a natural logarithmic function above $\SI{30}{\PE\per\nano\second}$, while the blue-dashed line serves as a reference for comparison with a slope of $1$. However, once the incident light intensity exceeds about $\SI{1000}{\PE\per\nano\second}$, the dynode signal also begins to lose linearity, compared with the blue dashed reference line. Moreover, after saturation, the dynode data points lie systematically above the reference line, indicating that the large amount of charge flowing through the voltage divider modifies the divider voltage and consequently the gain of the PMT. 

\section{Anode performance}\label{sec:anode}
\begin{figure}[htbp]
    \centering
    \includegraphics[height=0.38\linewidth]{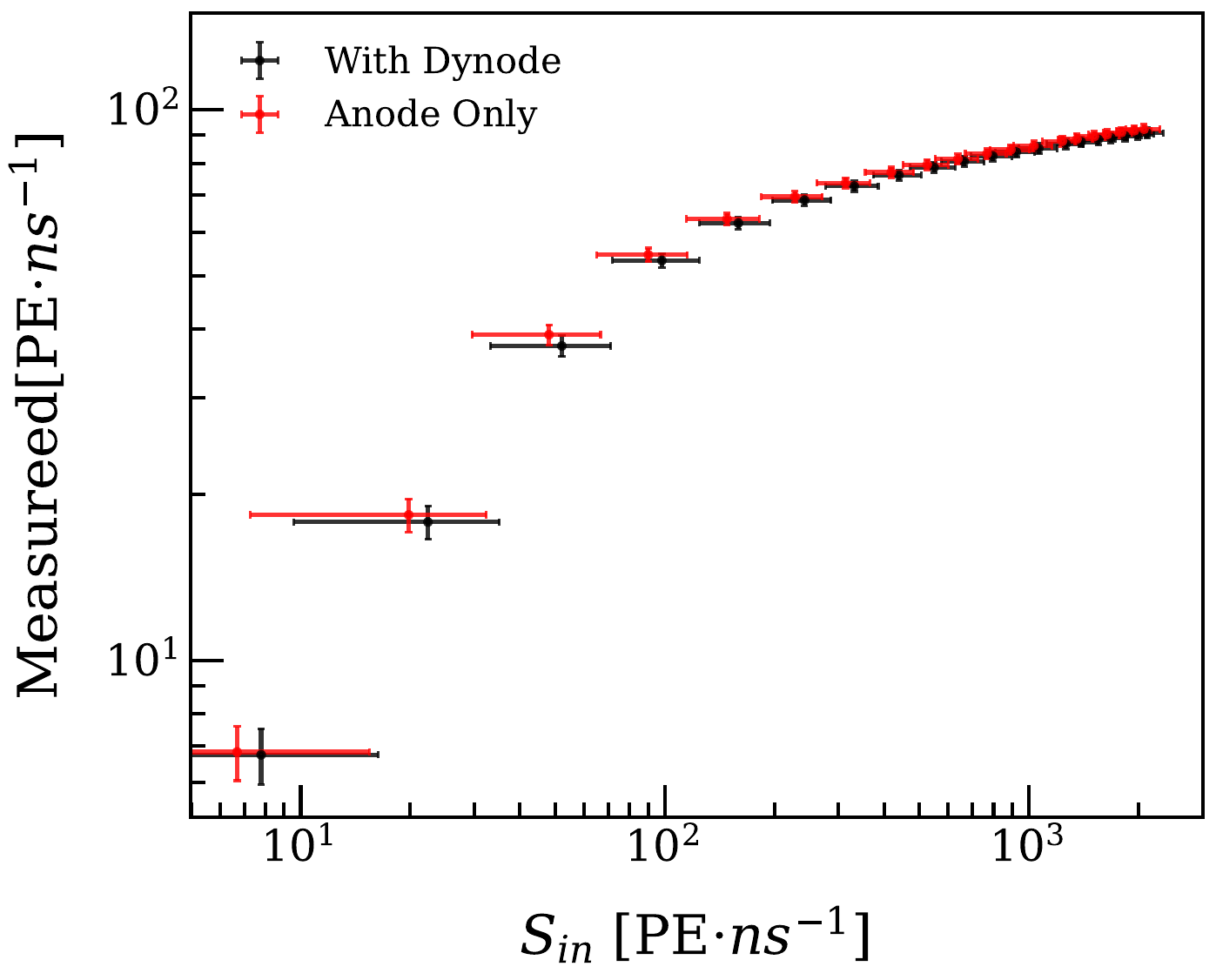}
    \includegraphics[height=0.38\linewidth]{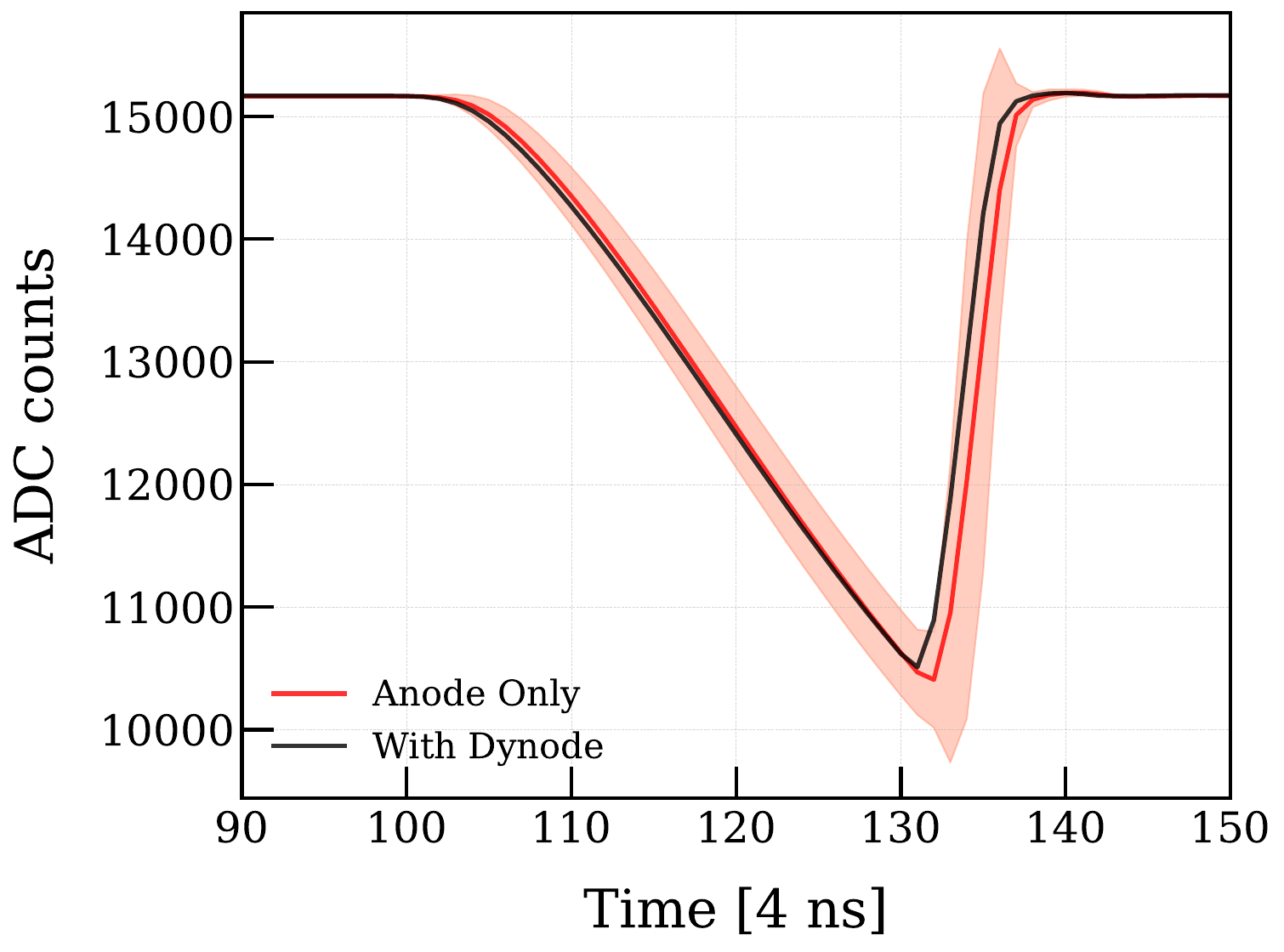}
    \caption{(Left) The readout response curve on the anode with (black) and without (red) the dynode-related components. 
    (Right) The raw waveform comparison of the anode only (red solid line, while the light red shadow region is the fluctuation), and the anode with dynode (black) under the light intensity of $21\,\text{PE} \cdot \text{ns}^{-1}$ from the LED. Anode response remains uncompromised because the dynode readout is electrically decoupled from the anode chain.}
    \label{fig:comparing}
\end{figure}

The anode signal of a PMT is the result of precedent multiplication of electrons, while the signal from the dynode is shaped by the induced charge of electrons drifting to the next dynode. In order to confirm that the design of the dynode readout does not compromise the regular performance of the anode for low energy \cevns\ detection:
\begin{itemize}[noitemsep, topsep=0pt, parsep=0pt, partopsep=0pt, label=\textemdash] 
\item we firstly compared the readout response and raw waveform with and without the dynode readout related components, as shown in figure~\ref{fig:comparing};
\item secondly, we chose the single photoelectron (SPE) response feature of a specific PMT to verify the design of the extended dynamic range base.
\end{itemize}
Since the S2 signal from \cevns\ falls in the range of $\SI{120}{\PE}$ to $\SI{300}{\PE}$ in RELICS, making the PMT dark counts contribution to the background relatively negligible, we do not discuss PMT dark current in this paper. 

\begin{figure}[htbp]
    \centering
    \includegraphics[width=9cm]{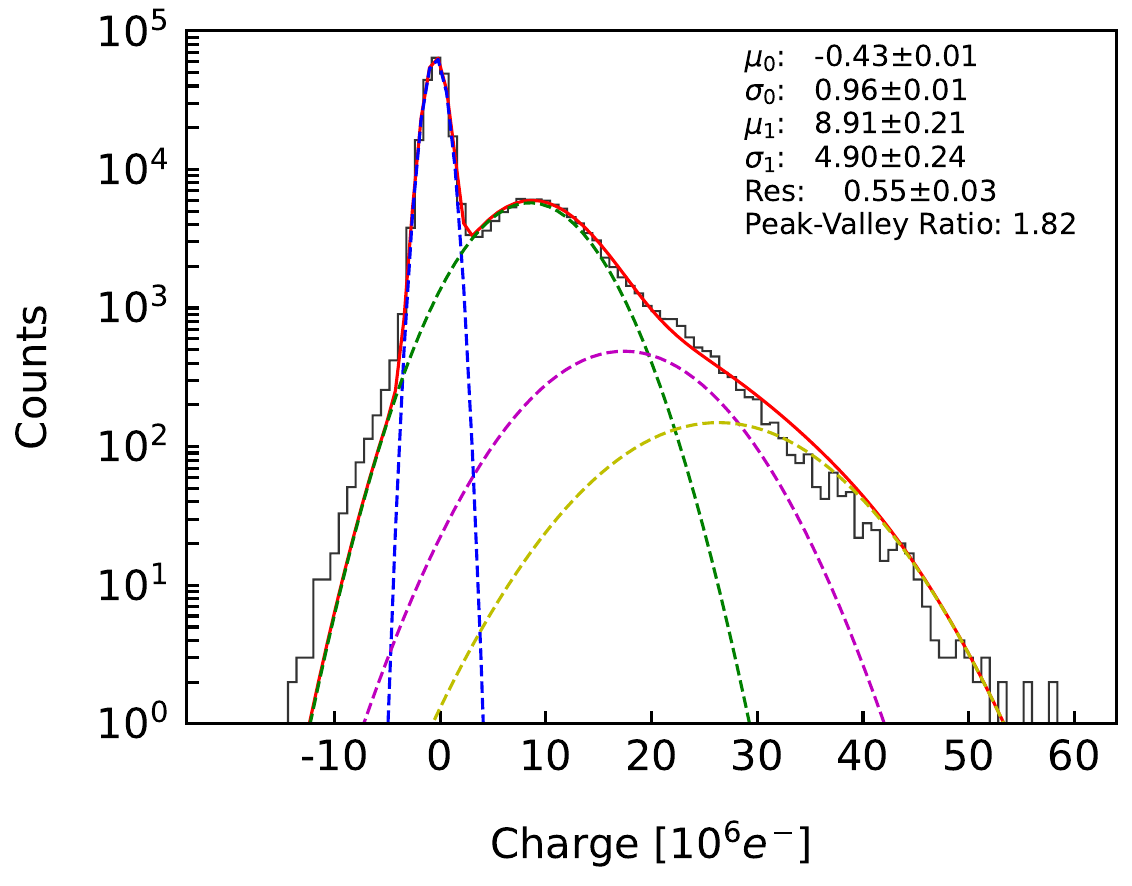}
    \caption{The PMT charge spectrum at $\SI{800}{\volt}$ and the corresponding fitting results using the model of eq.~(\ref{equ:spegain}) are shown. The blue dashed line is related to the baseline/pedestal, and the green dashed line is the spectrum of a single photoelectron. The magenta and yellow dashed lines indicate the double-photoelectron and triple-photoelectron distributions, respectively. The red solid line represents the combined fit. \textit{Res} is the SPE resolution. The overall mean and standard deviation are denoted as $\mu_1$ and $\sigma_1$. }
    \label{fig:spegain}  
\end{figure}

To achieve single photon emission, the LED was tuned to produce pulses shorter than $\SI{30}{\nano\second}$, with pulse rate set to fewer than $\mathcal{O}(10)$ pulses per hundred triggers~\cite{Huang:2014yfa}. The charge spectrum distribution is characterized by the LED under external trigger, as is shown in figure~\ref{fig:spegain}, fitted with a multi-Gaussian equation~\cite{Li:2015qhq}
\begin{equation}
    \begin{aligned}
        f(q) &= A_0 \cdot G(q, \mu_{0}, \sigma_{0}) + A_1 \cdot G(q, \mu_{1} + \mu_{0}, \sqrt{\sigma_{{1}}^{2} + \sigma_{0}^{2}})\;+ \\
            &\quad +\;A_2 \cdot G(q,2\mu_{1} + \mu_{0}, \sqrt{2\sigma_{1}^{2} + \sigma_{0}^{2}})\;+ \\
            &\quad +\;A_3 \cdot G(q, 3\mu_{1} + \mu_{0}, \sqrt{3\sigma_{1}^{2} + \sigma_{0}^{2}}),
        \label{equ:spegain}  
    \end{aligned}
\end{equation}
where $G(x, \mu, \sigma) = \mathrm{e}^{(x-\mu)^2/2\sigma^2}$ represents the standard Gaussian function centered at $x=\mu$ with standard deviation $\sigma$. The SPE gain, obtained from the fit, is $\SI{8.94}{\times 10^{6}}$, the corresponding SPE resolution (defined as the ratio of the standard deviation and mean value of the SPE spectrum) is $\SI{54.9}{\percent}$, with peak-to-valley ratio (the maximum height of the SPE spectrum divided by the minimum point between the SPE spectrum and the pedestal) is $1.84$, which shows that the SPE signal is clearly distinguishable from the noise. 

\section{Case study of impacts of muon signals in RELICS}\label{sec:largesig}
\subsection{Recovery time after muon S1}
\label{sec:recovery_after_S1}
Cosmic muons induce large signals in the TPC, which induce a light intensity as large as $\SI{500}{\PE\per\nano\second}$, shadowing the successful detection of the following signals, such as the \cevns. As the PMT base requires time to recharge the capacitors and recover from saturation, signals collected during this period are distorted, leading to inaccurate energy reconstruction. To mitigate this effect, we make further use of our setup for a specific case study for the muon S1 signal, aiming to quantitatively characterize the recovery process for typical cosmic muon S1 in RELICS. 

We evaluate the signal distortion caused by S1-induced saturation at varying energy levels on subsequent physical signals with differing temporal profiles. The test is performed by generating two pulses using an LED with a $\Delta t$ separation. The first signal mimics S1 of cosmic muons using a typical duration of cosmic muon S1 and a number of LED intensities. The second signal is weak and would not saturate PMTs, but is large enough to be measured accurately. By systematically scanning incident light intensity and delay time to the subsequent signal, we can quantitatively probe the time and energy dependencies of saturation recovery. 

\begin{figure}[htbp]
    \centering
    \includegraphics[width=8.6cm]{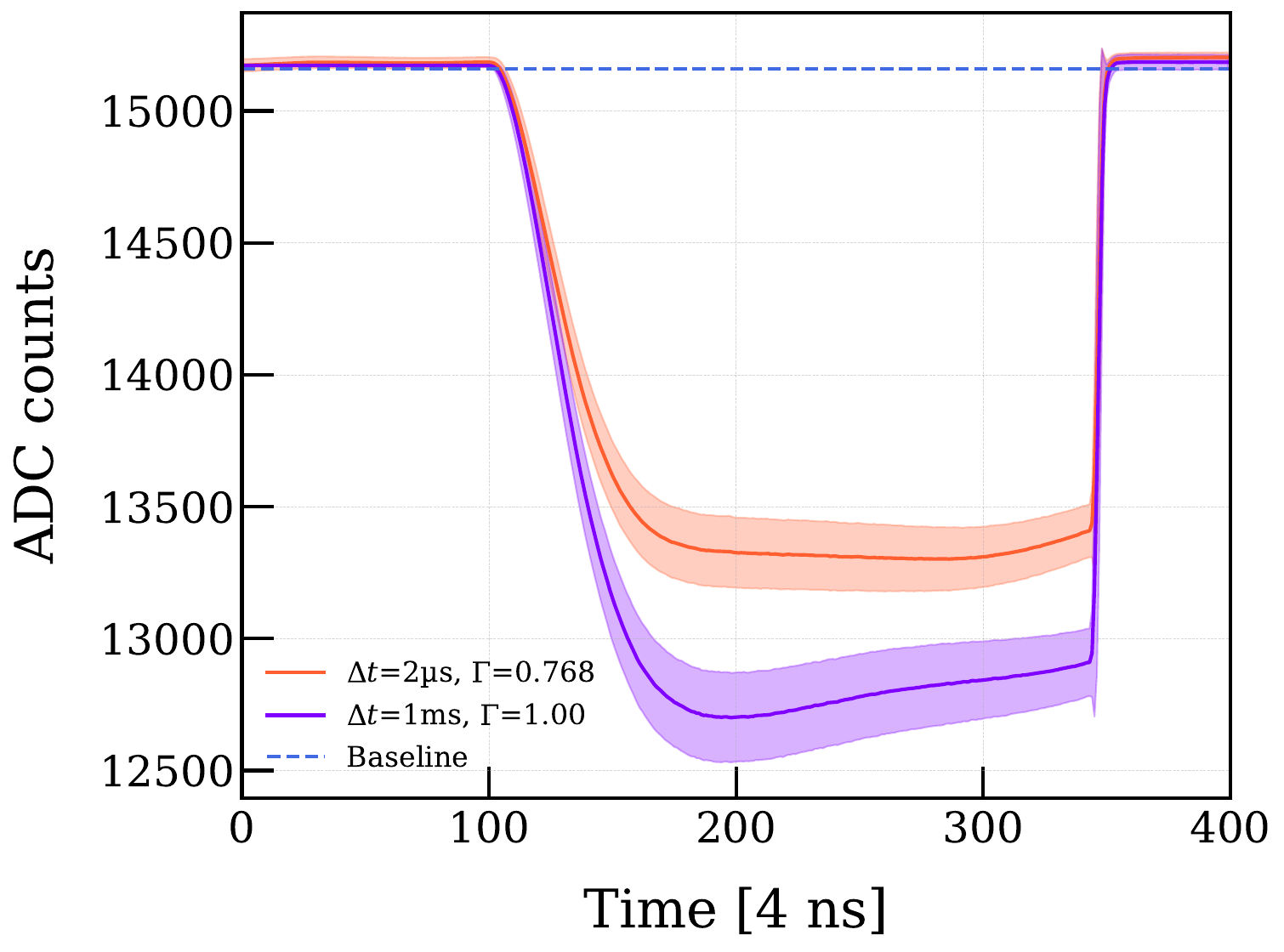}
    \caption{Example waveforms of the subsequent signal pulse with delay time at $\SI{2}{\micro\second}$ (light red) and $\SI{1}{\milli\second}$ (light purple) relative to the large intensity light, recorded at an incident light intensity of $\SI{13100}{\PE\per\nano\second}$. The waveforms represent the average value (solid) and variance (shadow) calculated from the raw data of each run. }
    \label{fig:wftimedelay}
\end{figure}

In order to quantify the effect of saturation, we define a dimensionless saturation correction factor $\Gamma$ to be the observed intensity divided by true intensity,
\begin{equation}
    \Gamma = \frac{S_{\text{obs}}}{S_{\text{true}}},
\end{equation}
where $S_{\text{true}}$ can be obtained with either the reference PMT or dynode readout. Both intensities are measured with pulse area (scaled to the unit of PE/ns under a fixed width of the LED drive pulse), as we care about the measured energy deposit in the subsequent pulse. As an example, figure~\ref{fig:wftimedelay} shows the waveform of the test signal with time delay of $\SI{5}{\micro\second}$ and $\SI{1}{\milli\second}$ from an extremely high incident light intensity, around $\SI{13100}{\PE\per\nano\second}$. 

\begin{figure}[htbp]
    \centering
    \includegraphics[height=0.34\textwidth]{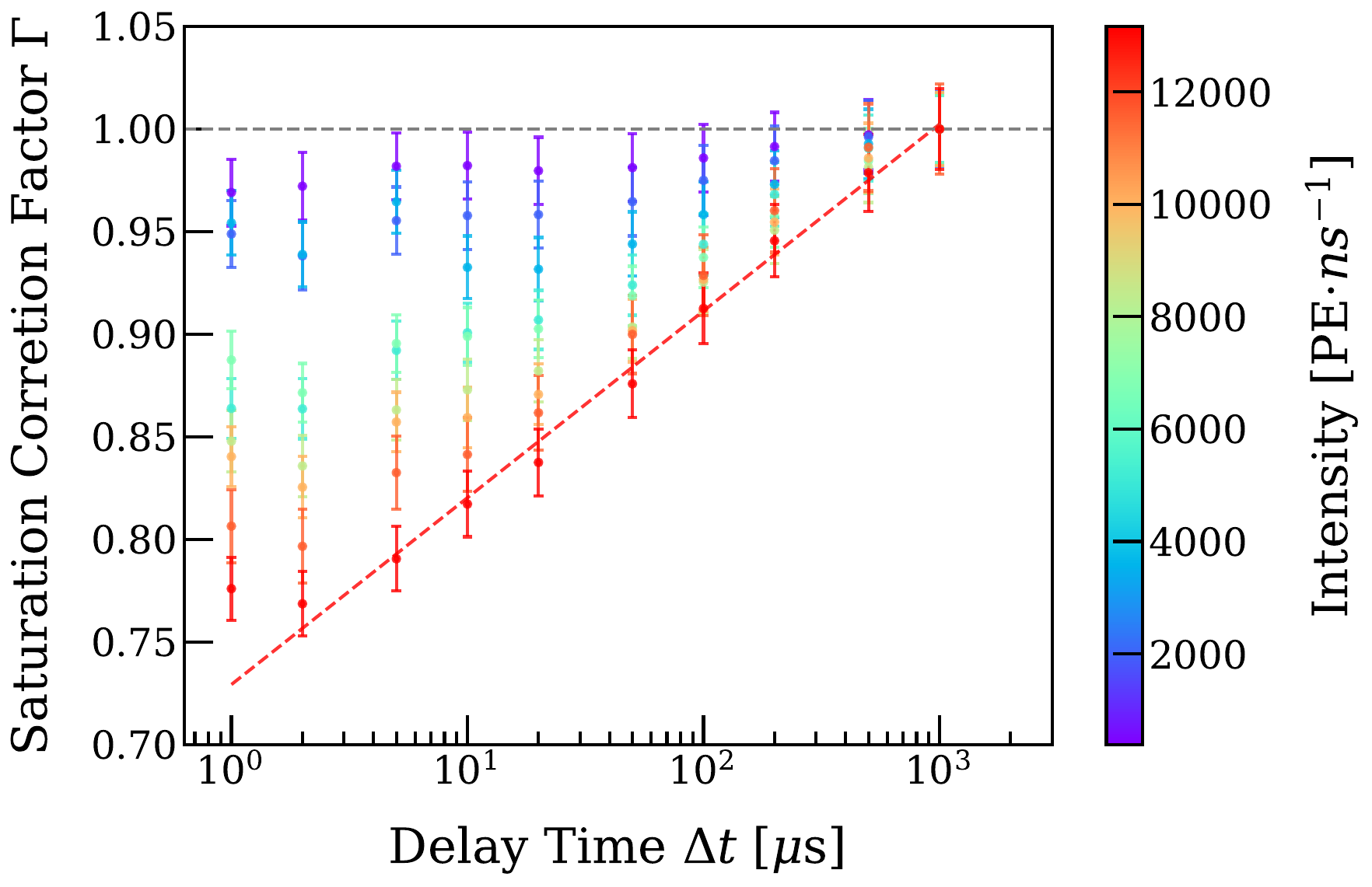}
    \hfill
    \includegraphics[height=0.34\textwidth]{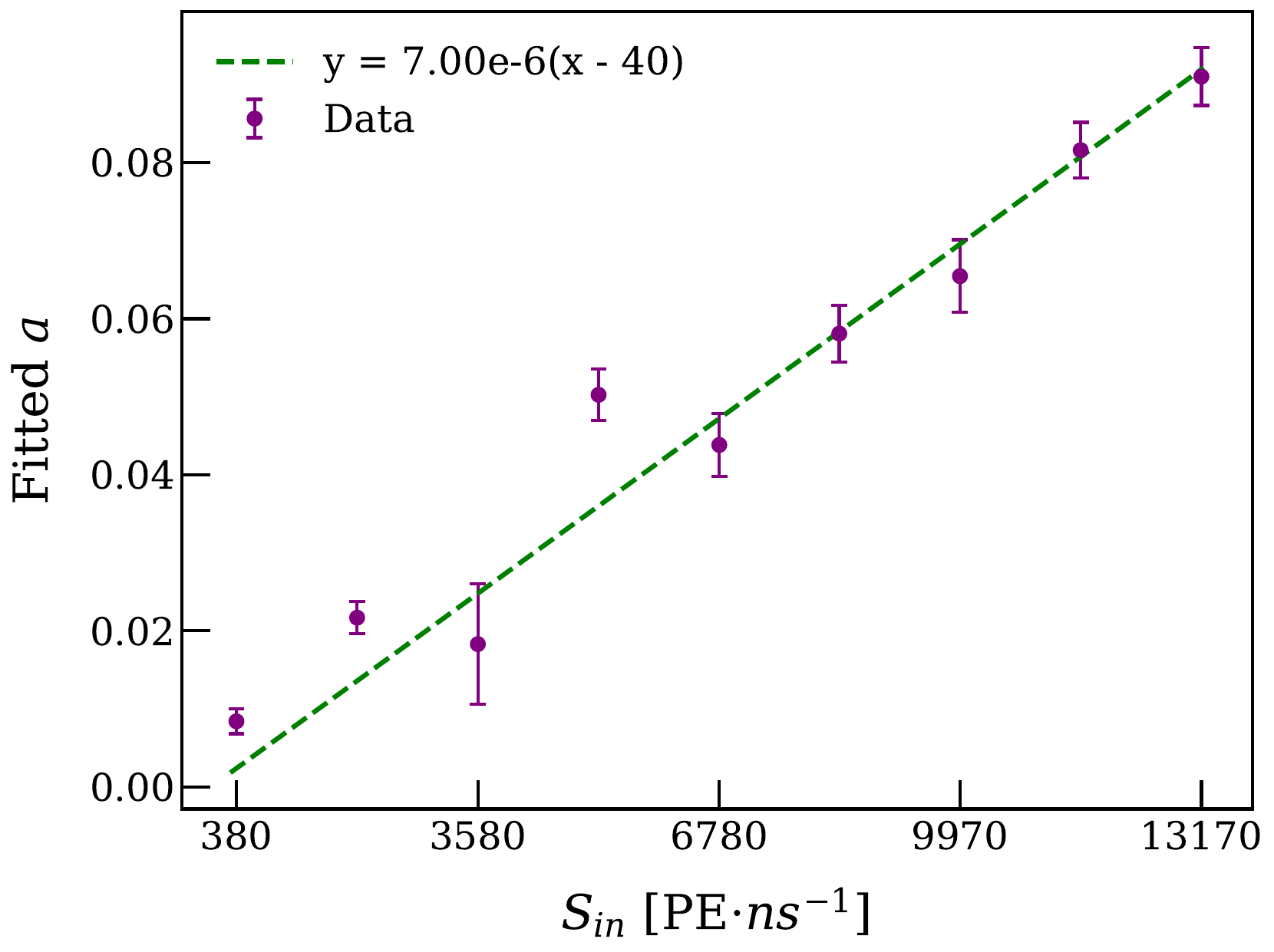}
    \caption{(Left) The saturation correction factor as a function of delay time, under different incident light intensities (values shown by colorbar), extrapolated from its relation with LED voltage, detailed in appendix~\ref{sec:appendix}. Exponential fit was applied in the $\SI{2}{\micro\second}\le\Delta t\le\SI{1}{\milli\second}$ region for each intensity. The red dashed line demonstrates the fitting results of the highest intensity using eq. (\ref{eq:surfact}). (Right) The saturation 2D dependence: parameter $a$ as a function of incident light intensity, which was converted from intensity measured by monitor PMT. The green dashed line was fitted using eq.~\eqref{eq:aslop}, while $S_{0}$ was fixed at $\SI{40}{\PE\per\nano\second}$ as discussed in the text. 
    }
    \label{fig:dependence_2D}  
\end{figure}

Scanning incident light intensity from $\SI{380}{\PE\per\nano\second}$ to $\SI{13100}{\PE\per\nano\second}$ and time separation from $\SI{1}{\micro\second}$ to $\SI{10}{\milli\second}$, we obtain the distribution of $\Gamma$ as a function of $\Delta t$, as shown in figure~\ref{fig:dependence_2D} (left). We assumed that the PMT saturation is fully recovered after a time delay of $\SI{1}{\milli\second}$, which was confirmed by data at $\Delta t = \SI{10}{\milli\second}$ (not shown). At the most extreme light intensity ($\sim\SI{13100}{\PE\per\nano\second}$), the lowest $\Gamma$ value is about $\SI{70}{\percent}$. This means that most of the subsequent signal will still be detected, while the energy might be distorted. It is worth noting that our muon-induced light signals in figure~\ref{fig:muon_duration} are mostly below $\SI{1000}{\PE\per\nano\second}$, corresponding to a minimum correction factor of $\SI{95}{\percent}$. 

To further characterize the dependence of PMT saturation on time and energy, we fit the results of the saturation correction factor in figure~\ref{fig:dependence_2D} (left) using the following equation:
\begin{equation}
    \Gamma(S_{\text{in}}, \Delta t)=  a(S_{\text{in}}) \log_{10}\left(\frac{\Delta t}{\SI{}{\micro\second}}\right) + \Gamma_{0}(S_{\text{in}}),\qquad \SI{2}{\micro\second}\le\Delta t\le\SI{1}{\milli\second},
    \label{eq:surfact}
\end{equation}
where $S_{\text{in}}$ is the intensity of incident light, measured in $\SI{}{\PE\per\nano\second}$. By calculating $a(S_{\text{in}})$ from figure~\ref{fig:dependence_2D} (left), we found that it varies linearly with intensity, as shown in figure~\ref{fig:dependence_2D} (right),
\begin{equation}
    a(S_{\text{in}}) = m\cdot (S_{\text{in}} - S_{0}),\qquad S_{\rm in}>S_0,
    \label{eq:aslop}
\end{equation}
where $S_0$ denotes the limit of anode saturation, which is fixed at $\SI{40}{\PE\per\nano\second}$ during fitting, and is dependent on the anode saturation point of figure~\ref{fig:dynamic_range}. The fitted parameter $m = \SI{7.00e-6}{\nano\second\per\PE}$ reveals that the linearity between PMT saturation and the part of incident light intensity exceeding the anode saturation point. 

Overall, the time and energy dependencies for a saturated anode can be expressed as
\begin{equation}
    \Gamma(S_{\text{in}},\Delta t) = 1 + m(S_{\text{in}} - S_{0}) \left[\log_{10}\left(\frac{\Delta t}{\SI{}{\micro\second}}\right) - 3\right],\qquad\SI{2}{\micro\second}\le\Delta t\le\SI{1}{\milli\second}.
\end{equation}
This function parameterizes the magnitude of the residual charge ratio of a signal after PMT saturation and should be used to quantify the fractional suppression of a subsequent low-energy signal due to the preceding high-energy muon interaction. In the analysis, this residual charge ratio is interpreted as a multiplicative attenuation factor describing the loss of observable signal amplitude. Accordingly, the \cevns\ signal amplitudes are corrected by applying the reciprocal of this factor, thereby compensating for the masking effect induced by the saturated muon pulse. This procedure restores the expected signal scale under the assumption that the distortion can be treated as a linear suppression of the charge response within the relevant time window after saturation.

\subsection{Waveform fidelity for muon S2}

\begin{figure}[htbp]
    \centering
    \includegraphics[width=9cm]{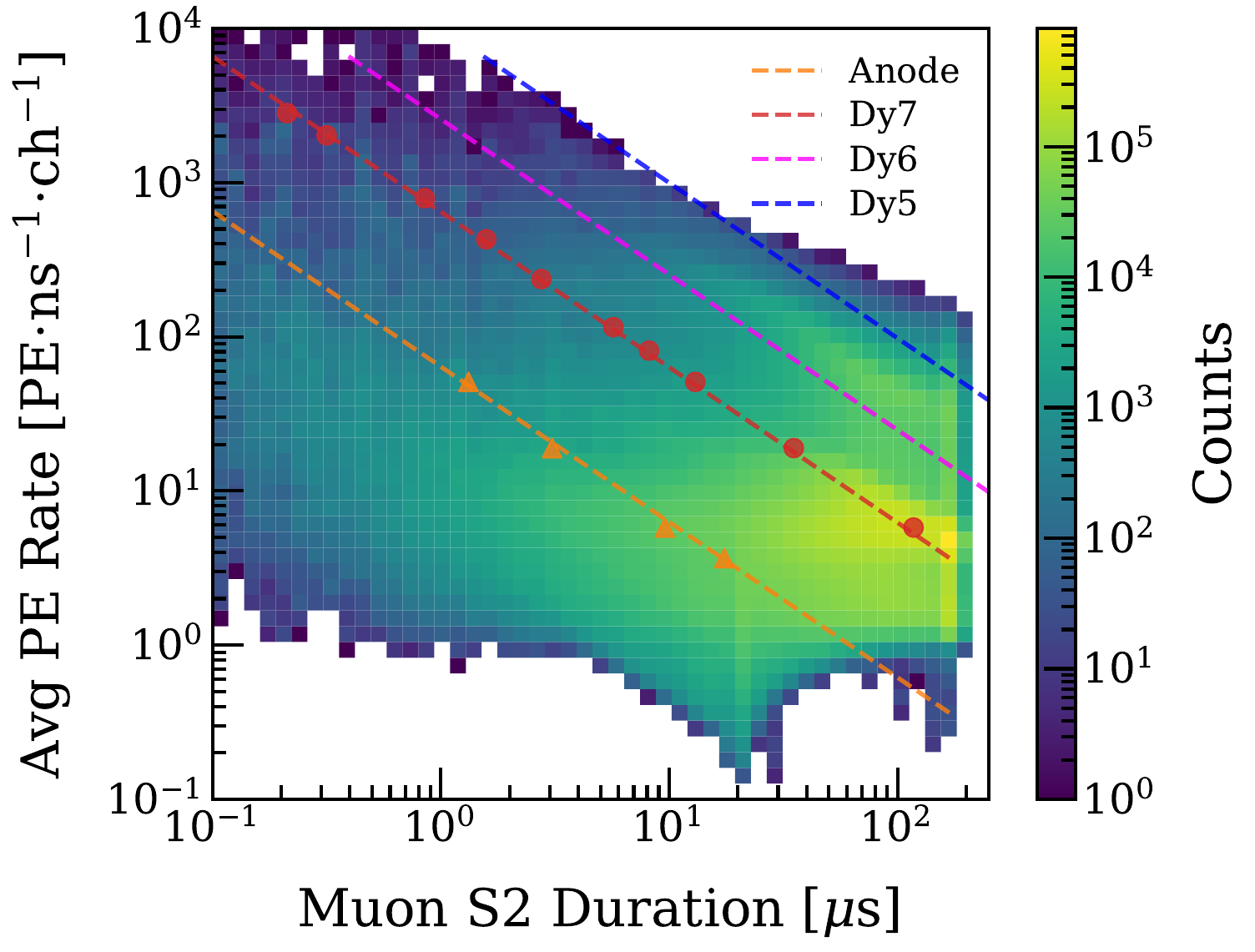}
    \caption{The 2D distribution for a simulated muon S2 per PMT~\cite{RELICS:2024opj}, in signal duration and average charge intensity. Overlay is the measured deformation time, where the pulse height is reduced to $\SI{50}{\percent}$ of the original, with a continuous signal at the intensity equal to the value on the $y$-axis. The red origin (orange triangle) represents the case of the 7th dynode (anode), fitted by a power-law function in the same color. The magenta dashed and blue dashed lines show the same lines with the sixth and fifth dynodes, respectively. }
    \label{fig:muon_duration}
\end{figure}

Compared to muon S1, the average light intensity of S2s induced by the muon is smaller. However, ionized electrons from a muon can occupy the entire drift time of $\SI{178}{\micro\second}$, such that the long-term effect on the PMT saturation cannot be neglected at the waveform level. As we rely on muon S2 to reconstruct muon tracks to suppress delayed electrons, a demonstration of the fidelity of a waveform under a large muon S2-like signal showcases the advantage of the dynode readout. Here, we tested the long-duration LED response of the PMTs at a constant light intensity over the entire drift time, sampling across intensities expected during a muon S2 in RELICS. In figure~\ref{fig:muon_duration}, we report the time waveform infidelity that happens as a consequence of saturation, defined as the time the output from the PMT drops to $\SI{50}{\percent}$ of the original height after a continuous light. Red circles and orange triangles denote the measurements with dynode and anode readout, respectively. Power-law functions were fitted to the data in the same color. These are overlaid with a population of muon S2 signals in signal duration and intensity, and the fraction of muon S2 waveforms that retain high fidelity is estimated by counting events below the fitted curves. We conclude that the current dynode-7 configuration covers $\SI{68}{\percent}$ of muon S2 signals, significantly higher than the $\SI{12}{\percent}$ from the anode. This shows that the dynode readout recovers 5 times more muon S2 signals than could be achieved with anode readout alone, maintaining superior energy information. Exploiting the observed linear relation in dynode-7 and anode, we extrapolate this trend to the sixth dynode (magenta dashed line) and the fifth dynode (blue dashed line), obtaining estimated coverages of $\SI{94}{\percent}$ and $\SI{98}{\percent}$, respectively.

\section{Discussion and conclusion}\label{sec:discussion}
The extended dynamic range base designed in this work mitigates the saturation of PMTs under intense light signals, significantly extending the linear response range. The extended dynamic range enables accurate detection across a wide spectrum of event energies. The dynode signal will be used for track reconstruction of high-energy cosmic muons, which will soon be demonstrated in the RELICS TPC and its prototypes. The saturation was investigated from both energy and timing perspectives, along with its impact on subsequent low-intensity light signals. Our bench test results indicate that PMT saturation caused by cosmic ray muon signals leads to less than $\SI{5}{\percent}$ distortion in subsequent physics signal measurements in the vast majority of cases, but the distortion can last for as long as $\SI{1}{\milli\second}$. These studies offer valuable references for correcting distortions caused by saturation effects in later signal processing. Furthermore, the extended dynamic range base proposed here has broad application potential for future LXe experiments. As the scale of LXe detectors continues to grow and detection sensitivity improves, balancing the high sensitivity required for low-energy signals with the linear response needed for high-energy signals remains a critical technical challenge. The dynamic base scheme provides an effective solution to this dilemma. 

Further, we showcased two essential PMT performances related to muon signals with our current design, related to the S1 and S2, respectively. In the first case with the high-intensity shadowing from muon S1, we demonstrated the fast recovery time of PMT for subsequent signals, like the \cevns. In the second case, we further demonstrated that by using the dynode-7 readout, we secured the fidelity of most of the muon S2s, enabling future endeavors to reconstruct muon tracks in RELICS TPC to suppress the dominating delayed electron background. We also demonstrated that the energy-time correlation of the S2 saturation point follows a clear power-law relation, such that the optimal dynode to use for de-saturation can be decided by simple scaling of a power-law function and calculating the muon signal coverage. Besides muon signals, we note that for the presumed axion-like particles~\cite{PhysRevD.109.075044, Mirzakhani:2025bqz, PandaX:2024sds, NEON:2024kwv}, which are produced in the reactor and observed in the RELICS detector with $\mathcal{O}(\SI{}{\mega\electronvolt})$ energy, can potentially benefit from the dynode signal. A dedicated optimization on the reactor axion signal is ongoing, following the same procedure detailed in this paper. 

In summary, the dual-channel readout board developed for the RELICS PMT array extends the linear dynamic range of a R8520-406 by nearly two orders of magnitude, maintaining linearity up to $\mathcal{O}(\SI{e3}{})~\SI{} {\PE\per\nano\second}$ on the dynode output while preserving the anode sensitivity required for sub-keV \cevns\ detection. Bench tests demonstrate that the highest-energy events in the RELICS detector will induce a $\SI{5}{\percent}$ loss of the energy of a subsequent event ($>\SI{5}{\micro\second}$) and fully recover at a time scale of $\SI{1}{\milli\second}$. We further parameterized their relation by a simple two-variable function of incident light intensity and delay time. These results imply that a surface-level LXe TPC can simultaneously accommodate low-energy recoil physics and high-rate muon monitoring without compromising energy reconstruction or event timing resolutions. Beyond RELICS, the same architecture furnishes a practical path toward direct, unsaturated readout of MeV-scale interactions --- an essential capability for studies ranging from neutrino-less double beta decay ($0\nu\beta\beta$) searches to reactor axion investigations.

\acknowledgments
RELICS is supported by grants from the National Key R\&D program from the Ministry of Science and Technology of China (No. 2021YFA1601600), Natural Science Foundation of China (Nos. 12275267, 12405129, 12375095, 12521007, 12250011), Beijing Natural Science Foundation (Nos. QY23088, QY25008), and the CUHK-Shenzhen University Development Fund (No. UDF01003491). We acknowledge CNNC Sanmen Nuclear Power Company for hosting RELICS. The work of JJY and RZL is supported by the Start-up Funding of Westlake University. The work of RZL is supported by a bursary from the Marshall Foundation, awarded through Jesus College Cambridge.

\bibliographystyle{JHEP}  
\bibliography{biblio.bib}

\appendix
\section{LED configuration}\label{sec:appendix}
\begin{figure}[htbp]
    \centering
    \includegraphics[width=11cm]{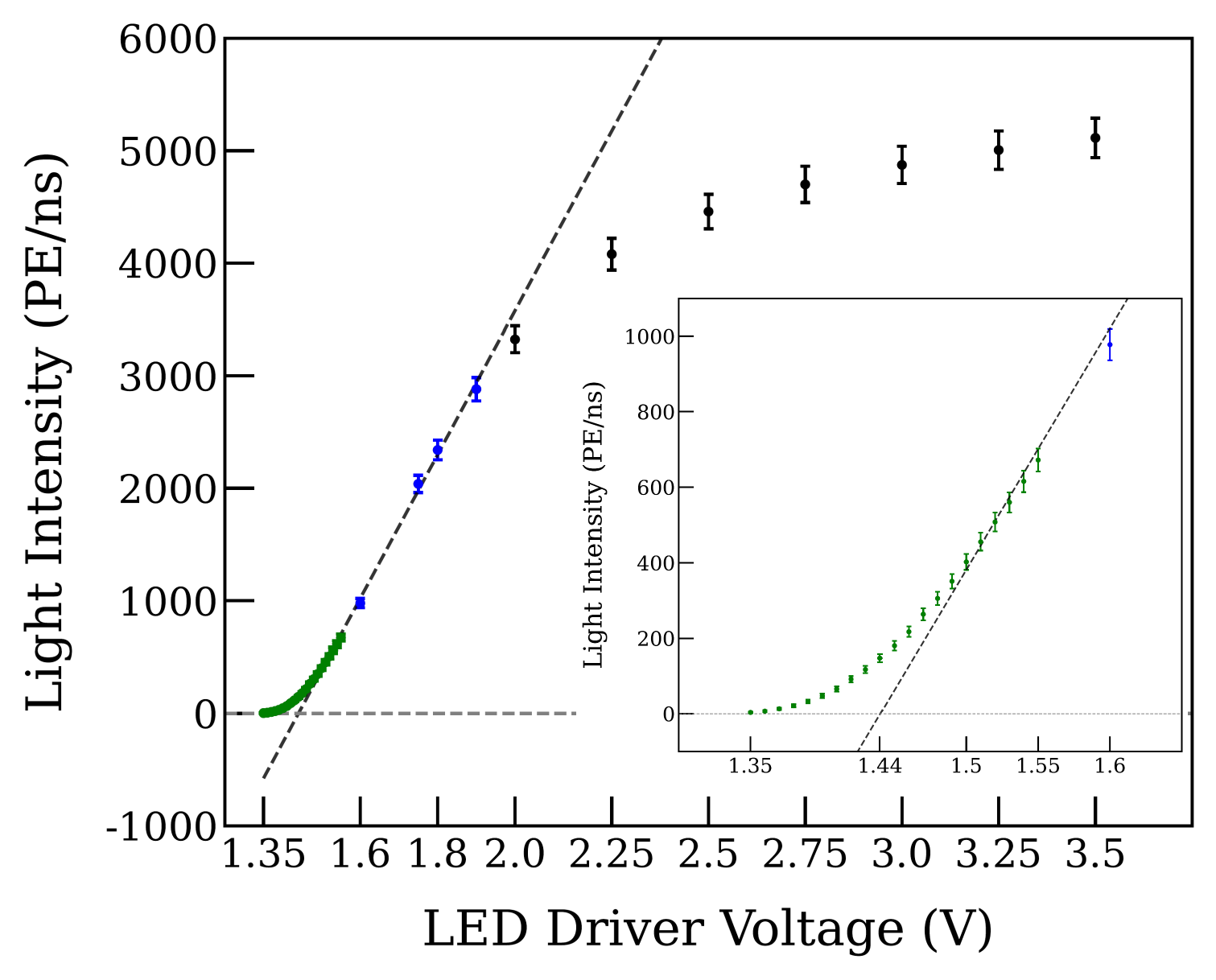}
    \caption{The LED configuration and PMT response mapping with fixed $\SI{150}{\nano\second}$ width of LED driven pulse. The $x$-axis is related to the driver voltage of the LED, and the $y$-axis is the light intensity received by the monitor PMT unit in $\SI{}{\PE\per\nano\second}$. The green data points related to the dynamic response range of the dynode, the dashed line was fitted with data between $\SI{1.6}{\volt}$ to $\SI{1.75}{\volt}$ (blue), which relates to the linear response readout range of the monitor PMT. Data points below $\SI{1.44}{\volt}$ are related to the spontaneous emission of the LED. We took the fitted results as the estimation value of light intensity beyond $\SI{2}{\volt}$ (black).}
    \label{fig:LED_test_range}  
\end{figure}

All LED-based measurements described in this appendix were performed using the setup shown in figure~\ref{fig:test_setup}. To test the dynamic readout range of the seventh dynode, we fixed the LED pulse width at $\SI{150}{\nano\second}$ and varied the driving voltage from $\SI{1.35}{\volt}$ to $\SI{2}{\volt}$, related from $\SI{5}{\PE\per\nano\second}$ to $\SI{3570}{\PE\per\nano\second}$, as is shown in figure~\ref{fig:LED_test_range}. This setup simulates the response and dynamic range of the dual-readout voltage divider for prompt scintillation S1 signals across different energy regions. Second, to study the saturation recovery time and its dependence on incident signal magnitude and duration, we fixed the LED pulse width at $\SI{150}{\nano\second}$ and adjusted the driving voltage from $\SI{1.5}{\volt}$ to $\SI{3.5}{\volt}$, which correspond to $\SI{380}{\PE\per\nano\second}$ to $\SI{13100}{\PE\per\nano\second}$, as shown in figure~\ref{fig:LED_test_range}, to simulate muon-excited S1 signals in LXe. Meanwhile, a separate test S2 signal with a $\SI{1}{\micro\second}$ pulse width and $\SI{5}{\PE\per\nano\second}$ intensity was used to mimic the S2 from \cevns. By varying the time interval between the test signal and the simulated muon signal, we scanned the saturation recovery time under different incident light intensities, which can be found in section~\ref{sec:recovery_after_S1}.

\section{Decoupler schematics}\label{sec:decoupler}
\begin{figure}[htbp]
    \centering
    \includegraphics[width=0.95\linewidth]{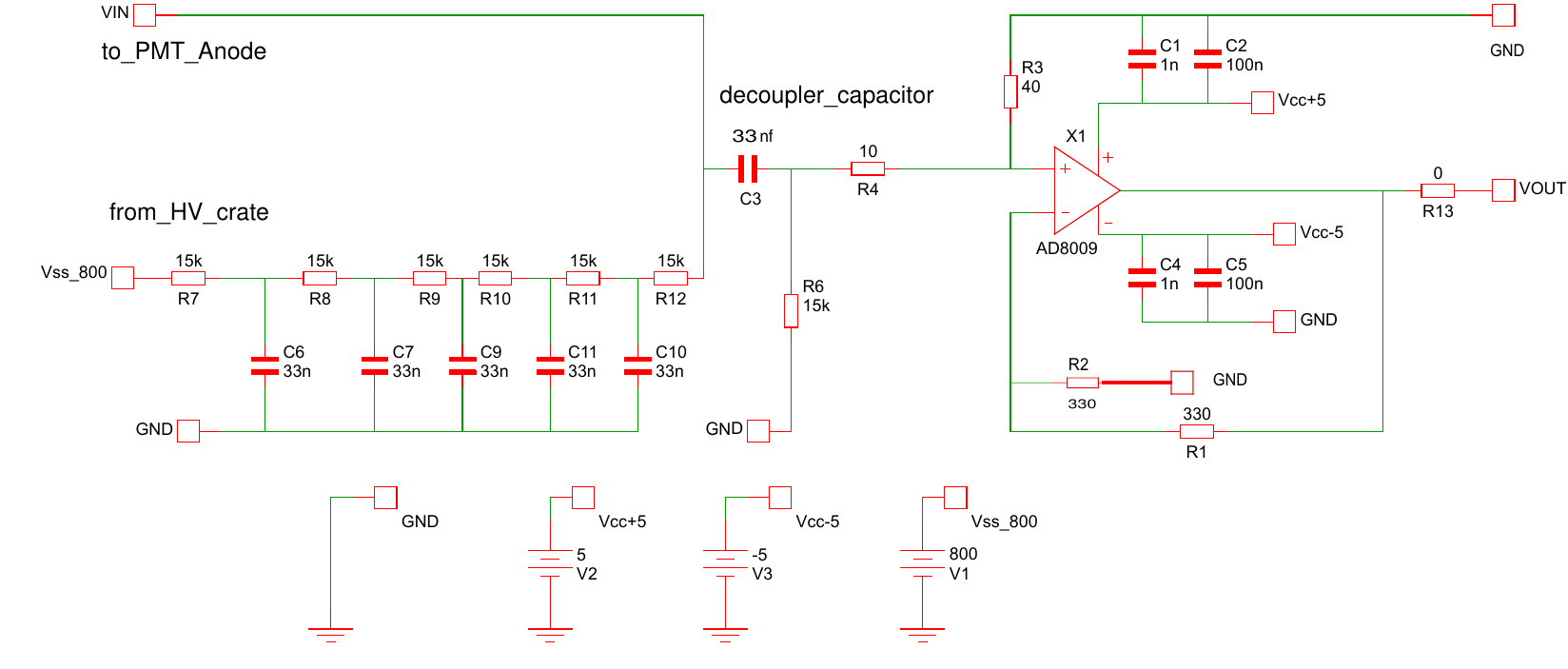}
    \caption{Decoupler circuit used in the RELICS experiment. The left section applies the high-voltage bias to the PMT anode through a five-stage RC low-pass filter that suppresses HV ripple and noise. Capacitor $C_{3}$ AC-couples the fast PMT signals to the readout while blocking the DC high voltage. The right part implements a unity-gain amplifier that protects the downstream electronics from transient currents generated by HV discharges in the TPC or connectors.
    }
    \label{fig:decoupler}  
\end{figure}

The decoupler circuit used in the RELICS experiment provides both high-voltage filtering and signal decoupling for the PMT readout. As shown in figure~\ref{fig:decoupler}, the high-voltage bias from the HV crate first passes through a five-stage cascaded RC low-pass filter, which suppresses HV ripple and noise. The AC-coupling capacitor $C_{3}$ then extracts the fast PMT pulses by blocking the DC high voltage and passing only the high-frequency components of the anode signal to the downstream electronics.
The right-hand part of the circuit consists of a unity-gain amplifier configured as a protection buffer. It is designed to intercept large transient currents that may be induced by sparking discharges inside the TPC or at HV connectors. In the event of such a discharge, the amplifier and its surrounding network will intercept the damaging voltage excursions from propagating into the DAQ system. Under normal operating conditions, the amplifier faithfully reproduces the PMT signal with a gain of one, providing a stable, low-impedance output for digitization.

\end{document}